\documentclass[letter,apj,twocolappendix,numberedappendix,appendixfloats]{emulateapj}

\usepackage{natbib}
\usepackage{amsmath}
\usepackage{color}
\usepackage{epstopdf}
\usepackage[multidot]{grffile}

\usepackage{graphicx}
\graphicspath{{figures/}}

\usepackage{hyperref}
\hypersetup{
            pdftitle={Towards a Network of Faint DA White Dwarfs as High-Precision Spectrophotometric Standards},
            pdfauthor={Narayan et al.}
            bookmarks=true,
            bookmarksnumbered=true,
            pdfpagelabels=true,
            colorlinks=true,
            anchorcolor=blue,
            citecolor=blue,
            linkcolor=blue
           }

\shorttitle{Faint High Precision Spectrophotometric Standards}
\shortauthors{Narayan et~al.}

\begin{document}
%%%%%%%%%%%%%%%%%%%%%%%%%%%%%%%%%%%%%%%%%%%%%%%%%%%%%%%%%%%%%%%%%%%%%%%%%%%%%%%%%%%

\title{Towards a Network of Faint DA White Dwarfs as High-Precision
  Spectrophotometric Standards}

\author{G. Narayan\altaffilmark{1}, T. Axelrod\altaffilmark{3}, J. B.
Holberg\altaffilmark{2}, T. Matheson\altaffilmark{1}, A. Saha\altaffilmark{1},
E. Olszewski\altaffilmark{3}, J. Claver\altaffilmark{1}, C. W.
Stubbs\altaffilmark{5, 6}, R. C. Bohlin\altaffilmark{4},
S.~Deustua\altaffilmark{4}, A. Rest\altaffilmark{4},  }

\altaffiltext{1}{National Optical Astronomy Observatory, 950 North Cherry Avenue, Tucson, AZ, 85719}
\altaffiltext{2}{University of Arizona, Lunar and Planetary Laboratory, 1629 East University Boulevard, Tucson, 85721}
\altaffiltext{3}{University of Arizona, Steward Observatory, 933 North Cherry Avenue, Tucson, 85721}
\altaffiltext{4}{Space Telescope Science Institute, 3700 San Martin Drive, Baltimore, MD 21218}
\altaffiltext{5}{Department of Physics, Harvard University, 17 Oxford Street, Cambridge, MA, 02138}
\altaffiltext{6}{Harvard-Smithsonian Center for Astrophysics, 60 Garden Street, Cambridge, MA, 02138}
\email{gnarayan@noao.edu}

\begin{abstract}
We present initial results from a program aimed at establishing a network of
hot DA white dwarfs to serve as spectrophotometric standards for present and
future wide-field surveys. These stars span the equatorial zone and are faint
enough to be conveniently observed throughout the year with large-aperture
telescopes. Spectra of these white dwarfs are analyzed to generate a
non-local-thermodynamic-equilibrium (NLTE) model atmosphere normalized to
\emph{HST} colors, including adjustments for wavelength-dependent interstellar
extinction. Once established, this standard star network will serve
ground-based observatories in both hemispheres as well as space-based
instrumentation from the UV to the near IR. We demonstrate the effectiveness of
this concept and show how two different approaches to the problem using
somewhat different assumptions produce equivalent results.  We discuss lessons
learned and the resulting corrective actions applied to our program.
\end{abstract}

%%%%%%%%%%%%%%%%%%%%%%%%%%%%%%%%%%%%%%%%%%%%%%%%%%%%%%%%%%%%%%%%%%%%%%%%%%%%%%%%%%%

\section{Introduction}\label{sec:intro}

Sub-percent global standardization of photometric calibration in astronomy
remains elusive.  Major ongoing and planned astronomical surveys (including
LSST,\footnote{\url{http://www.lsst.org}}
PanSTARRS,\footnote{\url{http://pan-starrs.ifa.hawaii.edu/public/}}
SDSS,\footnote{\url{http://www.sdss.org}} the Dark Energy Survey with
DECam,\footnote{\url{http://www.darkenergysurvey.org/}}
\emph{JWST},\footnote{\url{http://www.jwst.nasa.gov/}}
\emph{GALEX},\footnote{\url{http://www.galex.caltech.edu/}} and
\emph{WISE}\footnote{\url{http://wise.ssl.berkeley.edu/}}) will have to attain
band-to-band photometry and all-sky uniformity to better than 1\% to realize
their full scientific promise.  As a concrete example, consider the use of Type
Ia supernovae to probe the history of cosmic expansion and determine the
properties of dark energy. At present, photometric calibration issues dominate
the uncertainty budget on the equation of state of the dark energy, $w$
\citep[e.g., ][]{Sullivan11, Suzuki12, Betoule14}.  Weak lensing tomography with LSST also demands
sub-percent accuracies in color for reliable photometric redshift
determination (e.g., \citealt{Gorecki14} require a systematic uncertainty of
0.005 mag or less for their photometric redshift simulations).

Most of these surveys (as well as all post-\emph{Hubble Space Telescope (HST)}
visible-band astronomy in the foreseeable future) are or will be pursued from
the ground.  For calibration at ground-based facilities, one source of
uncertainty is the time-variable transmissivity of the atmosphere, with both
chromatic (Rayleigh scattering, Ozone absorption, Mie scattering, molecular
absorption, aerosol) as well as gray (cloud) terms. Attenuation from clouds,
water-vapor and aerosols can change rapidly and on differing angular scales and
amplitudes, and over all timescales \citep[e.g., ][]{Querel2014}. A variety of
methods are currently used to track and account for such effects including
monitoring with Light Detection and Ranging (LIDAR) techniques, using dual-band
Global Positioning System (GPS) metrology \citep[e.g., ][]{Li2014}, and
modeling the atmosphere (e.g., with \texttt{MODTRAN}).

There remain, however, key obstacles to obtaining fluxes/colors in physical
units at the 1\% level or better:

\begin{enumerate}
\item 
Vega was one of six A0V stars used to establish the color zero point on the
photometric system of \citet{johnsonmorgan1953} by defining the mean $U-B$ and
$B-V$ color of the six stars to be zero.  This definition was further extended
to Cousins $R_{C}-I_{C}$.  Vega's spectral energy distribution (SED) was tied
to tungsten-ribbon filament lamps and laboratory blackbody sources employed as
fundamental standards \citep[][and references therein]{oke70, Hayes75}.  The
laboratory sources (placed at distances of order a mile) and Vega were observed
through a telescope and spectral scanner, but corrections are necessary to
account for the extinction through the atmosphere along very different paths.
In particular, the wavelength-dependence of the aerosol scattering could not be
ascertained with sufficient confidence and the consequent uncertainty
permeates the empirical determination of Vega's SED.  Alternatively, stellar
atmosphere models of Vega have also been used as its intrinsic SED, in lieu of the
empirical comparison against a laboratory source (especially for extension into
the UV and IR).  Even if the models were perfect, there are other
complications.  Vega's SED is punctuated with several unusually shaped
absorption lines.  Vega has an excess of NIR emission longwards of
1--2 $\mu$m, likely a result
of its dust ring \citep{Bohlin14b} and possibly its  rapid rotation
\citep{Peterson2006}.  It also has an excess of UV emission relative to a 9400K
model (a result of its rapid rotation \citep{Bohlin14}). These may introduce
systematic errors when such models are used.  As a result, the absolute
calibration of Vega has known deficiencies at the $\sim 1$\% level in the
visible region and at the few percent level at near IR and near UV regions
\citep{blackwell83, selby83, mountain85}. 

\item 
Calibrations transferred from any one primary standard to other objects around
the sky using ground-based observations suffer from seasonal and site-specific
systematic errors especially in the gray component of extinction through the
terrestrial atmosphere.  Without a well-spaced network of calibrators,
``self-calibration'' \citep{schlafly12} or ``ubercal'' \citep{Padmanabhan08}
methods cannot guarantee systemic all sky uniformity.  They are especially vulnerable to any 
seasonal variations in declination dependence of atmospheric extinction.
\item
Other currently used primary SED standards, such as BD+17\arcdeg4708 and P330E,
are too bright to be directly imaged by the large aperture telescopes employed
by dark energy studies. Secondary standards are typically no fainter than
magnitude 15, which is not within the operating dynamic range of surveys such
as PanSTARRS, DES, and LSST.  Indirect links require bridging across $\sim
8$~mags of dynamic range to calibrate the surveys.  This allows various
systematic errors to creep in.
\end{enumerate} 

The best way to circumvent the issues with the terrestrial atmospheric
extinction would be to fly a standard laboratory source, such as National
Institute of Standards and Technology (NIST) calibrated standard, above the
atmosphere. The planning and execution of such an expensive experiment
takes a long time, however, and much progress can be made towards establishing
sub-percent accuracy spectrophotometry now using existing facilities and at
much lower cost.   

DA white dwarfs (WDs) have atmospheres dominated by hydrogen and thus are the
simplest stellar atmospheres to model. The opacities are known from first
principles and, in the temperature ranges (20,000--80,000K) in which we are
interested, photospheres are purely radiative and photometrically stable. The
SED from such an atmosphere can be defined by just two parameters:  effective
temperature ($T_{\text{eff}}$) and surface gravity ($\log g$ ).  These
parameters can be determined spectroscopically from a detailed analysis of the
\ion{H}{1} Balmer profiles, without reference to any photometry.  Thus the SED,
from the UV to the IR, can be calculated at arbitrary spectral resolution and
used to calibrate any photometric passband or spectroscopic system.  A flux
normalization in any chosen passband and accurate derivation of
reddening/extinction are the only other quantities required to fully
characterize the received flux and establish such objects as spectrophotometric
standards.  \citet{Bohlin14b} used the \emph{HST} Space Telescope Imaging
Spectrograph (STIS) to observe three bright DA WDs spanning a range of
temperatures. These objects are  unaffected by reddening as interstellar
absorption at Ly$\alpha$ implies a low \ion{H}{1} column density, corresponding to a
reddening of $E(B-V)<0.0005$. \citet{Bohlin14b} found their relative flux
distributions to be internally consistent with model predictions from
spectroscopic $T_{\text{eff}}$ and $\log g$ to better than 1\% in the
wavelength range $0.2 - 1.0\mu$m.  This level of internal consistency is
superior to direct comparison with Vega fluxes from \emph{HST}/STIS
\citep{Bohlin04}.

The \emph{HST} photometric scale is thus defined relative to {\it model} SEDs
for 3 nearby DA WDs \citep{Bohlin07}. We employ the \emph{HST} ``Vegamag''
photometric scale throughout this work.  \citet{Holberg06} used synthetic
photometry of DA WDs to place $UBVRI$, 2MASS $JHK$, SDSS $ugriz$ and Stromgren
$ubvy$ magnitudes on the \emph{HST} photometric scale to 1\%.
\citet{Holberg08} confirmed this calibration by using a set of DA WDs with good
trigonometric parallaxes that agreed at the 1\% level with their photometric
parallaxes from the Bergeron photometric
grid.\footnote{\url{http://www.astro.umontreal.ca/~bergeron/CoolingModels/}}

Modern large-telescope imaging surveys have an overlapping brightness
range where the signal-to-noise ratio ($S/N$) is excellent for $V \sim
17-19$ mag and where these objects are unsaturated in the survey data.
Thus we must extend the WD scale to fainter reference objects.

The three WDs used to define the CALSPEC calibration are bright and near enough
that line of sight extinction to them can be ignored. However, the WDs in our
desired brightness range are expected to have mild but easily characterizable
reddening that any experiment we design must be able to quantify. There may be
some faint DA WDs at 17 to 18 mag in the Galactic polar caps, but for all sky
coverage, we cannot get around having to deal with extinction. Notionally,
spectroscopy to derive $T_{\text{eff}}$ and $\log g$ of suitably chosen DA WDs
between 17 and 19 mag would be enough to define their SEDs.  However these
brightness levels place the WDs at sufficiently large distances where
interstellar extinction cannot be ignored.

Alternatively we could choose a purely empirical spectrophotometric
extension of the CALSPEC calibration by observing a more diverse class
of objects at the desired brightness with HST/STIS (so that the
measurements do not suffer from the effects of the terrestrial
atmosphere).  Such a sample could include redder objects and the
question of reddening would be inconsequential as we would measure the
net energy distribution.  However, to build a network of a couple of
dozen such stars, so that some subset of them are available at all
times from all sites though a variety airmasses, would require an
amount of observing time with STIS that would be prohibitively large.
It would also preclude the possibility of extrapolating the
calibration to wavelengths outside the range of the observations. The
STIS exposure time calculator (ETC) reports integration times of
several thousand seconds to reach a S/N of 100 for a 40,000K black
body with $V=18$ mag in just one of three configurations that would be
minimally required to cover the wavelength range from
$3000-18000$\AA.

The approach taken by us utilizes the principle behind the CALSPEC basis for
calibration, but builds up the system independently. By doing so, we also test
the proposition that the DA WD models are adequate, since we intercompare
results from many more objects for internal self-consistency. Our results need
to stand independently of any prior calibrations to be sufficiently robust,
which we think is worth the additional complications that our approach invokes.

In Table~\ref{tab:targets} we list 9 putative standard stars that are part of our Cycle 20
\emph{HST} Program (GO 12967, P.I.: Saha), including the best available
ground-based $V$-band magnitudes. Our aim in this program is to derive
effective temperature, surface gravity, and interstellar reddening for
these stars, that then fully 
parameterize (via the stellar atmosphere model) the SED incident upon the top of the Earth's atmosphere, 
which is normalized with respect to the \emph{HST} flux scale.

\begin{deluxetable*}{llcccc}
\tabletypesize{\scriptsize}
\tablewidth{0pt}
\tablecolumns{6}
\tablecaption{Primary DA White Dwarf Targets\label{tab:targets}}
\tablehead{
    \colhead{Object} &
    \colhead{Alternate ID} &
    \multicolumn{2}{c}{RA  (J2000)   Dec} &
    \colhead{$V$}                                  &
    \colhead{Spectral Observations\tablenotemark{a}}                     \\
    \colhead{}   &
    \colhead{}                                       &
    \colhead{{h}\phn{m}\phn{s}}                         &
    \colhead{\phn{\arcdeg}~\phn{\arcmin}~\phn{\arcsec}} &
    \colhead{(mag)}                                     &
    \colhead{} }
\startdata
SDSS-J010322.19-002047.7 &              & 01:03:22.1910 & $-$00:20:47.73 & 19.07 & G   \\
SDSS-J041053.63-063027.7 & WD-0408-066  & 04:10:53.6340 & $-$06:30:27.75 & 18.88 & GM  \\
WD-0554-165              &              & 05:57:01.3000 & $-$16:35:12.00 & 18.20 & GMI \\
SDSS-J102430.93-003207.0 &              & 10:24:30.9320 & $-$00:32:07.03 & 18.84 & G   \\
SDSS-J120650.40+020142.4 & WD-1204+023  & 12:06:50.4080 & $+$02:01:42.46 & 18.64 & G   \\
SDSS-J131445.05-031415.6 & WD-1312-029  & 13:14:45.0500 & $-$03:14:15.64 & 19.03 & GM  \\
SDSS-J163800.36+004717.7 & WD-1635+008  & 16:38:00.3600 & $+$00:47:17.80 & 18.82 & G   \\
SDSS-J203722.16-051303.0 & WD-2034-053  & 20:37:22.1670 & $-$05:13:03.03 & 18.91 & G   \\ 
SDSS-J232941.32+001107.8 & WD-2327-000  & 23:29:41.3250 & $+$00:11:07.80 & 18.58 & G   \\
\enddata
\tablenotetext{a}{References for Spectral Observations:\\
   G = Gemini-S + GMOS \\ 
   M = MMT + Blue Channel \\
   I = Magellan Baade + IMACS \\
}   
\end{deluxetable*}

The essential goal of this program is to establish a set of
spectrophotometrically self-consistent calibrated standard stars, tied as well
as possible to an absolute physical scale.  Along the way, we are empirically
testing the proposition that DA WD atmospheric models result in realistic SEDs
by examining whether observed flux ratios from stars with different atmospheric
parameters are consistent with model predictions, after derivation and
application of corrections to account for interstellar reddening. In this paper
we present initial findings that define the level of consistency it is possible
to achieve using these procedures.

We aim to place a select set of DA WDs between magnitude 17 and 19 on a common
physical scale for flux incident at the top of the terrestrial atmosphere. This
will remove uncertainties in currently used SED standards that are plagued
with uncertain corrections for extinction within the terrestrial atmosphere.
The expectation is that sub-percent internal consistencies can be achieved,
which will be an improvement by several factors over any currently employed
photometric or spectrophotometric system.  A set of thus better-calibrated
standard SED sources referred to a
physical (absolute) system, a subset of which is
always visible from any ground-based observatory, will serve as universal
standard calibration sources for spectrophotometry and photometry.

In \S~\ref{sec:obs} we describe the data attributes and reductions for both
the \emph{HST} imaging, as well as the ground based spectroscopy. In
\S~\ref{sec:phot} the photometry from the \emph{HST} imaging is described, along with
the procedure to put the measured magnitudes on the \emph{HST} photometric scale
defined by the three bright DA white dwarfs.  The modeling of the emergent and
received flux from each of the white dwarfs is described in
\S~\ref{sec:modeling}, as is the derivation of reddening/extinction to
reconcile model and observed fluxes. In \S~\ref{sec:alt} we present an
alternate approach to the reconciliation of model fluxes and observed counts
that bypasses some of the assumptions made in \S~\ref{sec:modeling} and show
that the two complementary approaches produce the same results to within the
accuracy limitations of the data at hand. This cross-validates some key
assumptions and results of both procedures.  In \S~\ref{sec:predicts} we
present predicted magnitudes of four stars in the respective passbands of three
current imaging surveys. We conclude (in \S~\ref{sec:Concl}) with a
discussion of caveats and lessons learned that have informed the observing
strategy for our \emph{HST} Cycle 22 program. 

\section{Observations and Data Reduction}
\label{sec:obs}

This program has three operational components. The first consists of
multi-band photometry from \emph{HST}/WFC3 images of each star that
accurately tie the fluxes to the \emph{HST} photometric scale. The
second is a grid of appropriate model atmospheres to generate
detailed, absolutely calibrated flux distributions and model spectra
(in particular the line profiles of Balmer lines) for each star.  The
third is high $S/N$ Balmer line spectra, nominally obtained with the
Gemini Multi-Object Spectrograph (GMOS) instrument on Gemini
South. The observed Balmer line profiles are analyzed in reference to
the model profiles to accurately determine 
$T_{\text{eff}}$ and $\log g$ for each star.
The model SEDs for stars with atmospheric parameters so determined are
then compared to the observed \emph{HST}/WFC3 photometry in multiple
bands, modulo a to-be-determined contribution from interstellar
reddening along the line of sight to each star. The best match between
model and measurement determines the reddening/extinction
parameter(s).

\subsection{WFC3 Observations}
\begin{figure}[htb]
\centering
\epsscale{1.2}
\plotone{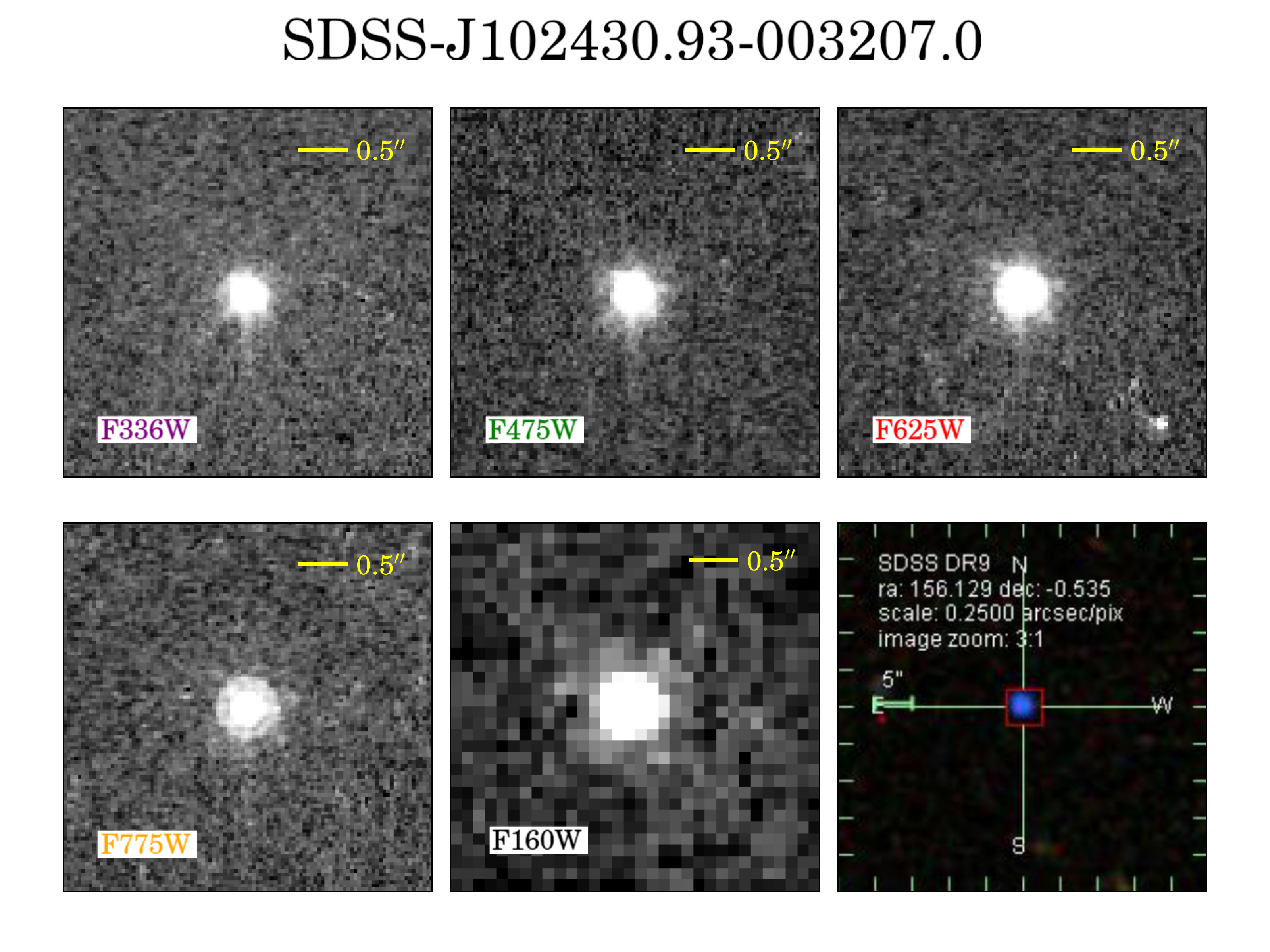}
\caption{WFC3 images of SDSS-J102430.93 and an SDSS finder. Exposure times have
been chosen to ensure high $S/N$ in each of our five filters. Observations
consist of 2--3 sub-exposures. Any observation with only 2 sub-exposures
will be supplemented with an additional Cycle 22 observation. Additionally, we
will obtain $F275W$ observations of each target in Cycle 22.} 
\label{fig:field} 
\end{figure}

Wide Field Camera 3 (WFC3) observations have been obtained for all of the 9
target stars listed in Table~\ref{tab:targets}, These observations yield
absolute photometric fluxes in the five bands ($F336W$, $F475W$, $F625W$,
$F775W$, and $F160W$) that are used to normalize the reddening corrected model
spectrum that defines the stellar flux distributions.  In Figure~\ref{fig:field}
we show the five WFC3 images for one of our targets, SDSS-J102430.93, along
with an SDSS finder chart image of this star.

All targets were placed near the center of the WFC3 UVIS2
detector. Each object is observed two or three times in each band with
small dithers to facilitate cosmic ray rejection. The bands were
chosen to ensure that the temperature, surface gravity, and reddening
could be independently determined, without significant
covariance. Stromgren $u$ ($F336W$) is chosen because it sits entirely
shortward of the Balmer jump and so allows a way to measure it.  For
our targets, the Balmer jump is sensitive to temperature. This
dependency, however, is different from that of the Paschen continuum
slope. The $F336W$ observations therefore help to mitigate
degeneracies between reddening and temperature---a critical
prerequisite for our study.  $F475W$, $F625W$, and $F775W$ emulate the
Sloan $gri$ bands, respectively, and were chosen so that, in addition
to measuring the Paschen continuum, they also provide a direct measure
in bands being used in large surveys today (e.g., SDSS, PanSTARRS, and
DES) and those planned in the future (LSST).  $F160W$ is the
\emph{HST}/WFC3 equivalent for the $H$ band, and it anchors
measurements of our objects as far into the infra-red as \emph{HST}
will allow.

The exposure times were chosen to allow for $S/N$ better than 200 for each of
the targeted white dwarfs while staying shorter than half of the time to
saturation. The arrangements of the various exposures times within an orbit
were crafted to minimize ``dead time'' between exposures and optimize program
efficiency. A post-flash flux was added to ensure the background level reached
12 electrons to mitigate charge transfer efficiency (CTE) losses. Additional
parallel observations that include stars within a few arc-minutes from each of
our target white dwarfs were obtained with the Advanced Camera for Surveys
(ACS).  Photometry for them, and an evaluation of their usefulness as
supplementary standard stars will be presented in a future paper.

\begin{figure*}[htb]
\centering
\plotone{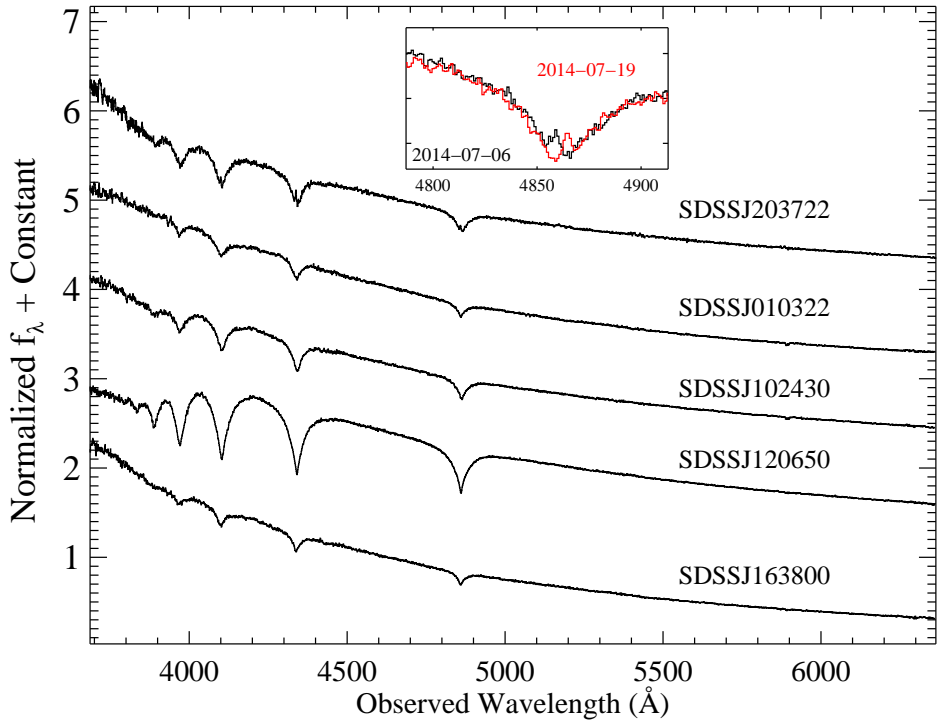}
\caption{GMOS spectra of 5 of the 9 targets for which we have complete
spectroscopic and photometric data. SDSS-J203722.16 has emission lines in the
cores of the Balmer lines; a magnified view of the H$\beta$ line is shown in
the inset for 2 epochs 13 days apart during which time the emission feature has
moved by 5.3 pixels. As the observations were
obtained under good seeing with a wide slit, the wavelength shift need not be solely due
to a change in the relative velocity of the emission feature.  Our spectra have $S/N$ between
80--170 with a dispersion of 0.92~\AA/pix. Even the weak H$\zeta$ feature is
clearly resolved.}
\label{fig:allspectra}
\end{figure*}

\begin{figure*}[htb]
\centering
\epsscale{1.2}
\plotone{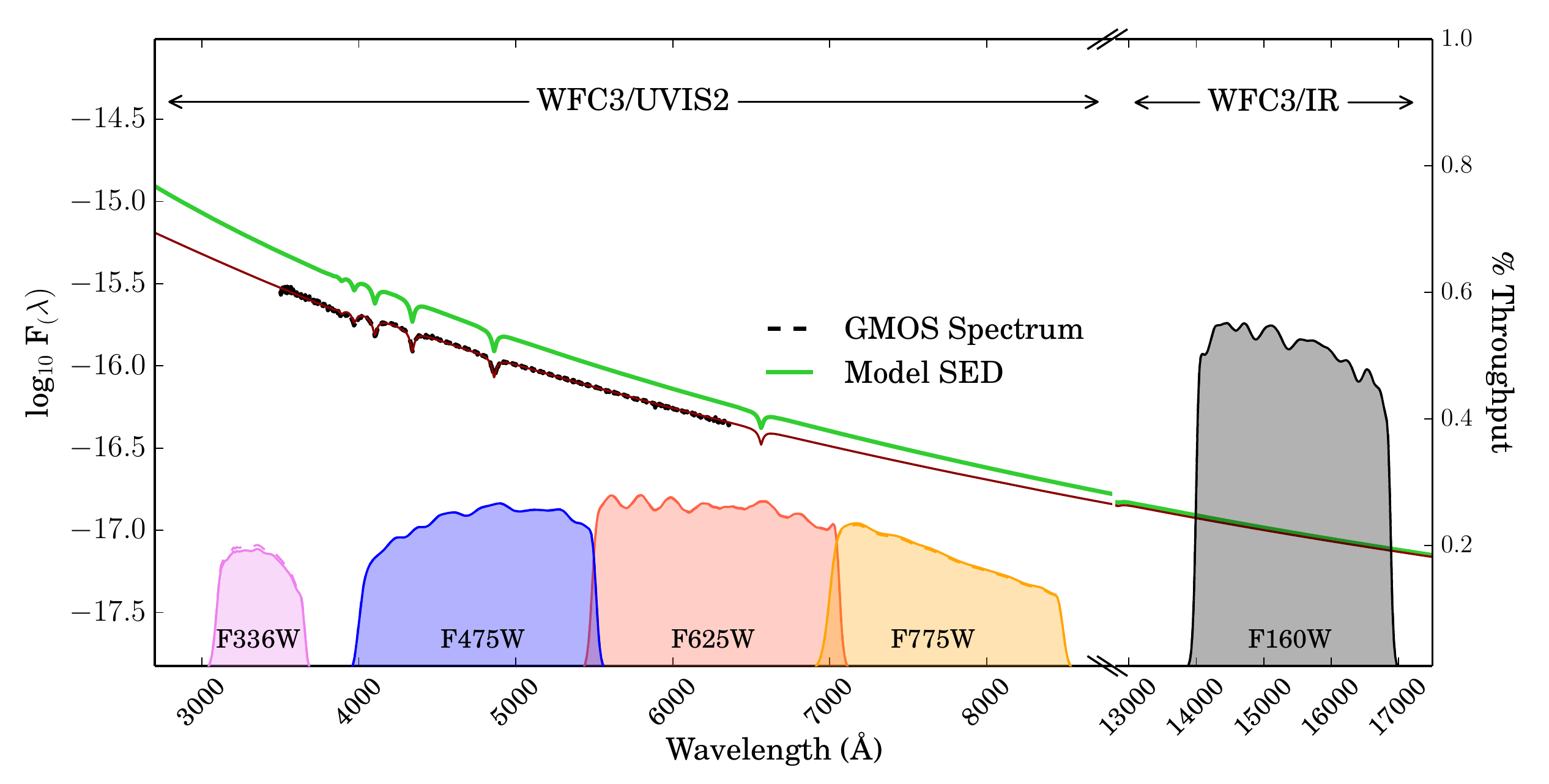}
\caption{GMOS spectrum (black) along with extinguished (red) and unreddened
(green) models of SDSS-J102430.93.  The transmissions of the various WFC3
passbands used in our program are shown in the shaded regions. The continuum
flux is removed from the spectral features before the Balmer profiles are fit.
The fit model with extinction does provide an excellent match to the observed
spectrum indicating that our spectroscopic flux calibration is reliable. }
\label{fig:spectra}
\end{figure*}

\subsection{GMOS Observations}
We used GMOS \citep{Hook04} at Gemini South to obtain moderate resolution
($R\sim1000$), high-$S/N$ spectroscopy of our WDs. The analysis of the detailed
shape of the \ion{H}{1} Balmer profiles (H$\beta$ to H$\zeta$) provides primary
spectroscopic estimates of $T_{\text{eff}}$ and $\log g$ that are independent
of photometry or SED slopes.

The observations were obtained in queue-scheduled mode on GMOS with the B600
grating and a long slit having a width of 1\farcs5.  Night sky lines had a
full-width at half maximum of $\sim$12\AA. In general, there were $6 \times
1500$s exposures made of each target. The inter-chip gaps between the three
GMOS CCDs were filled by taking three observations at one grating tilt and
three at another.  The entire combined spectrum contiguously covers the
wavelength region between 3500~\AA\ to 6360~\AA\ with dispersion of
0.92~\AA/pix. We resampled the spectra onto a single linear scale at 1\AA\ per
bin. We used standard IRAF\footnote{IRAF is distributed by the National Optical
Astronomy Observatory, which is operated by AURA under cooperative agreement
with the NSF.} routines to process the CCD data and optimally extract
\citep{Horne86} the spectra.  The overall $S/N$ per resolution element is very
high, averaging from 80 to 170 for our targets.  We used our own IDL routines
\citep{Matheson08} to flux calibrate the spectra using Feige 110
\citep{Stone77}.  Strictly speaking, the flux calibration is not needed (and in
any case rests on a less accurate basis than the ones we seek to establish),
and the analysis presented in \S\ref{sec:linefit} is independent of the flux
calibration of the spectra.  Nevertheless, getting the continuum slopes as
correct as possible helps when analyzing the shapes of the wide Balmer lines.

At this time, there are five targets (of the nine in Table~\ref{tab:targets})
for which we have complete spectra and photometry. Their spectra are shown in
Figure~\ref{fig:allspectra}. One of these, SDSS-J203722.16, exhibits narrow
emission lines in the cores of the Balmer absorption features.  The emission
lines shift with time. This could indicate the presence of a low-luminosity
companion, such as an M star, although there are other possibilities.  We
exclude this object from all further analysis. The remaining four spectra are
very characteristic of DA white dwarfs and appear free of contamination by
companions.

Because the calibration standard star was not always observed contemporaneously
with the targets, we validated the calibration by comparisons with the expected
fluxes obtained by preliminary reddened model fluxes. In Figure~\ref{fig:spectra}
we show such a comparison. As discussed in \S\ref{sec:modeling}, the method
used to analyze individual Balmer profiles is relatively insensitive to any
residual calibration uncertainties.  

The overall response of the grating and detector yielded a count rate spectrum
that peaks very near H$\beta$ and falls off significantly to the blue. At
H$\zeta$ the signal is approximately 30\% of peak, leading to significant $S/N$
gradient across the Balmer lines.  Fortunately the overall $S/N$ is very high,
averaging from 80 to 170 per bin for our targets.  For some targets, we have
obtained supplemental spectroscopic observations using IMACS \citep{Dressler11}
at Magellan, and the Blue Channel \citep{Schmidt89} at the MMT.  They have
similar $S/N$ and resolution, and cover the same spectral range as the GMOS
configuration.

\section{Photometric Analysis}\label{sec:phot}

As this program requires extremely accurate and precise determinations
of the white dwarf flux, we have elected to use two independent
methods to photometer our WFC3 observations. The drizzled
(\texttt{drz}) images from the Mikulski Archive for Space Telescopes
(MAST) are processed using \texttt{SExtractor} \citep{bertin96}. The
images are first masked to remove pixels in the \texttt{flt} image file  
that are only derived from a
single exposure (determined from the \texttt{ctx}
extension) and sources above the saturation threshold. A 64-pixel
background mesh is used with an 8-pixel filter size to smooth the
image and estimate the background locally. This procedure removes
virtually all spurious objects (cosmic rays, drizzling artifacts,
etc.). We use a range of apertures from 4--25 pixels to measure the
flux of each source. As WFC3 has a physical image scale of
0\farcs04/pix on UVIS and 0\farcs13/pix on the IR channel, this
corresponds to 0\farcs16--1\farcs0 on UVIS and 0\farcs52--3\farcs25 on the
IR channel. We construct curves of growth for each image from objects
with stellar point spread functions (PSFs). We use a simple filtering
routine to remove any objects that are more than 3$\sigma$ away from
the mean curve of growth in more than 3 apertures.  This removes any
objects with close companions and any non-stellar sources.  We use the
mean curve of growth of the image to interpolate between the measured
aperture flux in each pixel to determine the precise flux with a
physical aperture size of radius 0\farcs4. The weight \texttt{wht} map output from
\texttt{multidrizzle} is used to estimate the
uncertainties.

Measurements were also made using an interactive program written by
one of us (AS).  A curve of growth (COG) as function of aperture size
is constructed and the background level is set to be that value for
which the COG is flat in the radius range 9 to 12 pixels
(0\farcs36--0\farcs48) in the UVIS images and between 6 and 8 pixels
radius (0\farcs78--1\farcs04) in the IR channel data.  The `background'
thus measured is actually within the low level wings of the stellar
PSF, but allows the use of a tight aperture so that $S/N$ is not
compromised and chances of encountering any unremoved cosmic rays or 
other pathologies is minimized. Using published encircled energy
curves,\footnote{\url{http://www.stsci.edu/hst/wfc3/phot\_zp\_lbn}}
measurements made this way can be corrected systematically for
``infinite'' radius apertures (and referred to the true background),
which include all of the incident light from a given object. Look-up
tables for these corrections were made for each passband and applied
as appropriate.  These interactively made measurements, although
time-consuming, are judged to be a good investment towards quality
assessment because they quickly reveal the common kinds of image
pathologies that defy assumptions inherent in our data models.  These
interactive measurements were made on both \texttt{flt} and
\texttt{drz} images, where the \texttt{flt} images were corrected for
variations in pixel area. Concordant measurements on both sets of
images indicate whether the \texttt{multidrizzle} process compromises
data quality (e.g., by being overzealous about CR rejection and
cropping the tips of bright stars). Agreement across the two measuring
methods establishes that methodology-dependent systematic errors are
not being injected.

\begin{figure*}[htb]
\centering
\epsscale{1.2}
\plotone{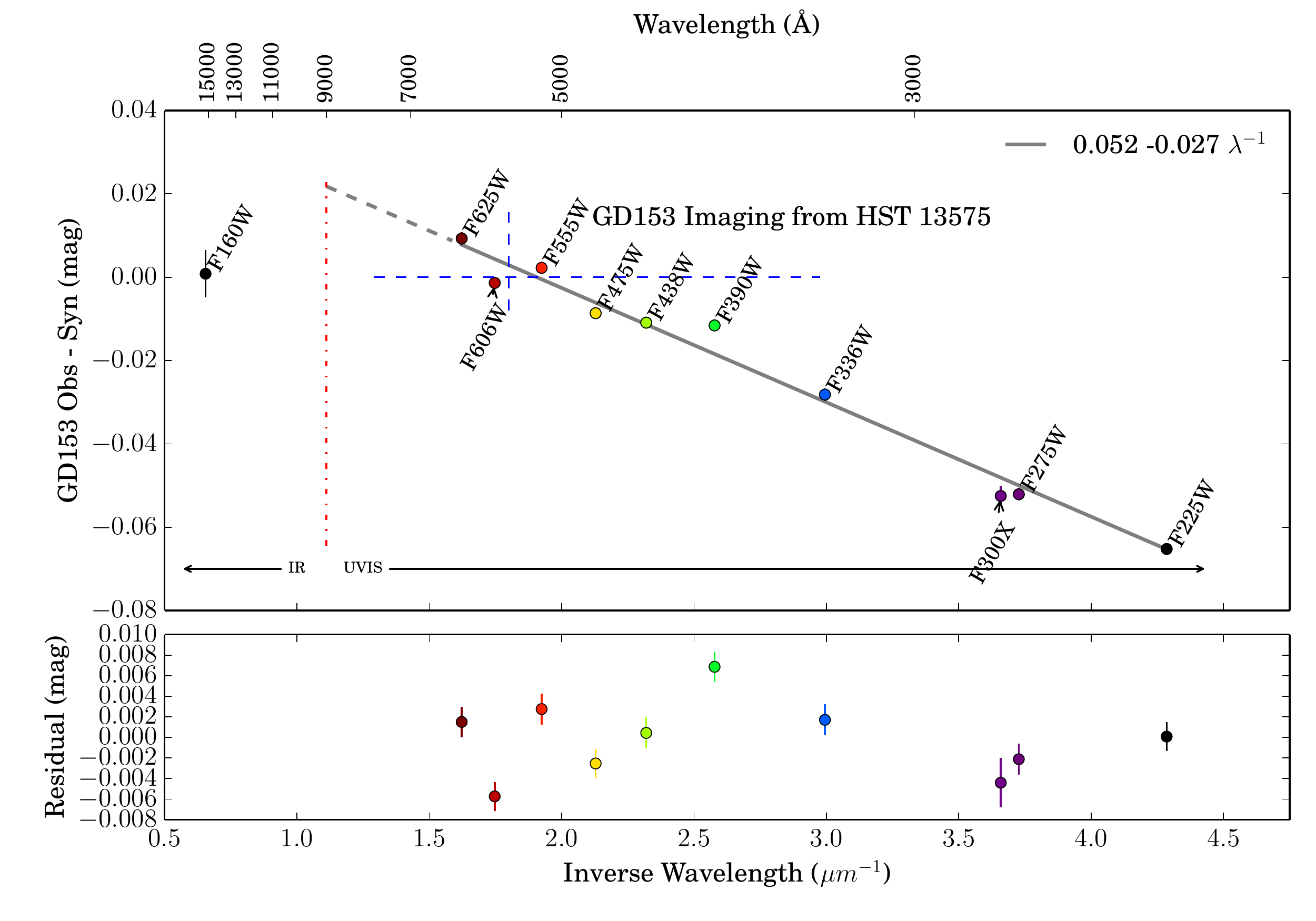}
\caption{(Top) Differences between measured magnitudes with fiducial
  MAST zeropoints and synthetic CALSPEC magnitudes for GD153
  (\emph{HST} program GO 13575, P.I. S. Deustua). We found a strong
  linear trend with frequency, indicating a difference between the
  zeropoints reported by MAST (which are the same for UVIS1 and
  UVIS2), and the true zeropoint. (Bottom) Residuals to the linear fit
  to the data are typically below 0.005mag. The trend with frequency
  suggests that this effect is likely a result of differing quantum
  efficiencies of the two UVIS detectors. This finding was
  independently verified by STScI staff scientists and will be
  corrected with new flat-field calibrations applied to future
  \emph{HST} data products.}
\label{fig:gd153resid}
\end{figure*}

As the WFC3 detector sensitivity is a function of time, the zeropoints are time
dependent. Rather than rely on the fiducial zeropoints supplied by MAST, we
measured the zeropoints directly using contemporaneous observations of CALSPEC
primary standard, GD153, obtained through \emph{HST} GO program 13575 (P.I.: S.
Deustua). These are applied as an additive adjustment to the fiducial MAST
zeropoint. Subrastered images of GD153 are available with the target at the
same position on the detector as our DA white dwarfs, however not in all of our
passbands ($F775W$ is missing). We are obtaining contemporaneous observations of all three
primary standards in Cycle 22 to determine instrumental zeropoints, and tie our
measurements directly to this photometric system.

In this work, we have measured GD153's instrumental magnitudes using the same
process for our targets, as described above. We applied the fiducial MAST
zeropoints to our instrumental magnitude, and compared these magnitudes against
synthetic photometry of the CALSPEC model of GD153
(\texttt{gd153\_fos\_003})\footnote{\url{ftp://ftp.stsci.edu/cdbs/calspec/gd153\_fos\_003.fits}}
through the WFC3 passbands.  The synthetic photometry is described in
\S\ref{sec:modeling}, by Equation~\ref{eqn:synflux}. While this procedure is
not as optimal as tying our measurements directly to the three \emph{HST}
primary standards, it is best methodology possible given the extant data.

We found a difference between the synthetic and measured magnitudes that
exhibits a strong linear trend with passband frequency. The measurements in each
passband, and the linear trend is shown in Figure~\ref{fig:gd153resid}. There
is no process through which our measurement technique, which treats all the
images individually, can introduce such a linear trend with frequency. The
largest discrepancy was found in the ultraviolet and the smallest differences
in the optical. No significant difference was found in the WFC3/IR channel,
which has a completely different detector and light path through the camera.
The observed trend therefore indicated a possible difference between the
reported and true efficiency of the UVIS detectors.  

\begin{deluxetable*}{cccccccccc}[!htbp]
\tabletypesize{\scriptsize}
\tablewidth{0pt}
\tablecolumns{10}
\tablecaption{DA White Dwarf Observed Magnitudes, Fit Parameters, and Synthetic Magnitudes\label{tab:pho}}
\tablehead{
    \colhead{Object} &
    \colhead{Passband} & 
    \colhead{Observed} & 
    \colhead{$A_{V}$} & 
    \colhead{$T_{\text{eff}}$} & 
    \colhead{$\log g$} & 
    \colhead{Synthetic} & 
    \colhead{Extinction} & 
    \colhead{Extinguished} & 
    \colhead{Residual} \\
    \colhead{Name} &
    \colhead{} & 
    \colhead{Magnitude\tablenotemark{a}} & 
    \colhead{} & 
    \colhead{} & 
    \colhead{} & 
    \colhead{Magnitude} & 
    \colhead{($R_{V}=3.1$)} & 
    \colhead{Magnitude} & 
    \colhead{} \\
    \colhead{} &
    \colhead{} & 
    \colhead{(mag)} & 
    \colhead{(mag)} & 
    \colhead{(Kelvin)} & 
    \colhead{} & 
    \colhead{(mag)} & 
    \colhead{(mag)} & 
    \colhead{(mag)} & 
    \colhead{(mag)} \\
    \colhead{(1)} &
    \colhead{(2)} & 
    \colhead{(3)} & 
    \colhead{(4)} & 
    \colhead{(5)} & 
    \colhead{(6)} & 
    \colhead{(7)} & 
    \colhead{(8)} & 
    \colhead{(9)} & 
    \colhead{(10)} \\
}
\startdata
        & $F336W$ &  17.344 (0.003) &        &        &      & 17.176 (0.001) &  0.171 & 17.347 (0.008) & $-$0.003\\
        & $F475W$ &  19.189 (0.003) &        &        &      & 19.058 (3E-4) &  0.125 & 19.183 (0.006) & $+$0.006\\
SDSS-J010322.19   & $F625W$ &  19.424 (0.003) &  0.101 &  59530 &  7.53 & 19.340 (4E-5) &  0.088 & 19.428 (0.004) & $-$0.004\\
        & $F775W$ &  19.588 (0.004) & (0.005) & (260) & (0.02) & 19.522 (7E-5) &  0.068 & 19.590 (0.003) & $-$0.002\\
        & $F160W$ &  20.105 (0.005) &        &        &      & 20.081 (7E-5) &  0.021 & 20.102 (0.001) & $+$0.004\\
\\
        & $F336W$ &  17.315 (0.004) &        &        &      & 16.856 (7E-4) &  0.453 & 17.309 (0.008) & $+$0.006\\
        & $F475W$ &  19.004 (0.003) &        &        &      & 18.679 (1E-4)&  0.331 & 19.010 (0.006) & $-$0.006\\
SDSS-J102430.93   & $F625W$ &  19.170 (0.003) &  0.269 &  40620 &  7.75 & 18.937 (1E-4) &  0.234 & 19.171 (0.004) & $-$0.002\\
        & $F775W$ &  19.291 (0.003) & (0.005) & (124) & (0.01) & 19.111 (2E-4) &  0.180 & 19.291 (0.003) & $-$0.001\\
        & $F160W$ &  19.718 (0.005) &        &        &      & 19.658 (2E-4) &  0.055 & 19.712 (0.001) & $+$0.005\\
\\
        & $F336W$ &  17.301 (0.003) &        &        &      & 17.213 (0.002) &  0.089 & 17.302 (0.013) & $-$0.000\\
        & $F475W$ &  18.776 (0.002) &        &        &      & 18.708 (4E-4) &  0.064 & 18.773 (0.009) & $+$0.003\\
SDSS-J120650.4  & $F625W$ &  18.912 (0.002) &  0.052 &  23650 &  7.89 & 18.871 (1E-4) &  0.046 & 18.916 (0.006) & $-$0.004\\
        & $F775W$ &  19.033 (0.003) & (0.008) & (46) & (0.01) & 19.000 (2E-4) &  0.035 & 19.035 (0.004) & $-$0.002\\
        & $F160W$ &  19.443 (0.004) &        &        &      & 19.428 (2E-4) &  0.011 & 19.439 (0.001) & $+$0.004\\
\\
        & $F336W$ &  17.113 (0.003) &        &        &      & 16.778 (0.001) &  0.337 & 17.115 (0.008) & $-$0.002\\
        & $F475W$ &  18.925 (0.002) &        &        &      & 18.676 (2E-4) &  0.247 & 18.923 (0.006) & $+$0.002\\
SDSS-J163800.36   & $F625W$ &  19.136 (0.003) &  0.200 &  64610 &  7.43 & 18.961 (5E-5) &  0.174 & 19.135 (0.004) & $+$0.001\\
        & $F775W$ &  19.279 (0.003) & (0.005) & (425) & (0.02) & 19.144 (9E-5) &  0.134 & 19.278 (0.002) & $+$0.001\\
        & $F160W$ &  19.743 (0.004) &        &        &      & 19.704 (9E-5) &  0.041 & 19.745 (0.001) & $-$0.002\\
\enddata
\tablenotetext{a}{All magnitudes are reported in the HST VEGAmag system.
Uncertainties are quoted in parentheses}
\end{deluxetable*}

This finding was independently replicated by staff scientists at
STScI. As the two WFC3/UVIS CCDs were manufactured on different wafers
and foundry production runs, their physical properties, such as
quantum efficiency and thickness, differ.  Studies at STScI have found
that UVIS2 is $\sim30\%$ more sensitive in the UV than UVIS1.  In the
range $3500-7000$\AA, the two CCDs have comparable sensitivity, while
at wavelengths longer than 7000\AA, UVIS1 is a few percent more
sensitive. Furthermore, the two detectors have different detection
layer thicknesses.  For UVIS1 the median thickness is $16.04 \pm 0.23$
microns, and for UVIS2 it is $15.42 \pm 0.58$ microns\footnote{\url{http://www.stsci.edu/~INS/2010CalWorkshop/wong.pdf}}.

New flat-fields that treat the two UVIS detectors individually have been
created and will be applied to future MAST products. For this work, we correct
the fiducial MAST zeropoints for each of our passbands with the linear relation
we found. These corrected zeropoints are applied to the measured flux to
produce the observed magnitudes for the targets used in this work. The
photometry and measurement uncertainties for these targets is listed in column
3 of Table~\ref{tab:pho}. The measured magnitudes are therefore described by

\begin{equation}\label{eqn:zp1}
\begin{split}
m^{O}_{T,i} &= -2.5\log_{10}(\phi^{O}_{T,i}) + \text{ZP}^{\text{MAST}}_{T} + \Delta\text{ZP}^{\text{GD153}}_{T}
\end{split}
\end{equation} 

\noindent where $m$ is the natural magnitude, $\phi$ the measured flux of the
object, $i$, ZP$^{\text{MAST}}_{T}$ is the instrumental zero point published in
MAST, and $\Delta$ZP$^{\text{GD153}}_{T}$ is the correction to the fiducial
MAST zeropoint derived, as described above, using observations and
synthetically predicted photometry of GD153, observed through passband $T$.
``O'' is used to denote observed quantities.

\section{Modeling}\label{sec:modeling}

\subsection{Line Profile Fitting}\label{sec:linefit}

\begin{figure*}[htb]
\centering
\epsscale{1.2}
\plotone{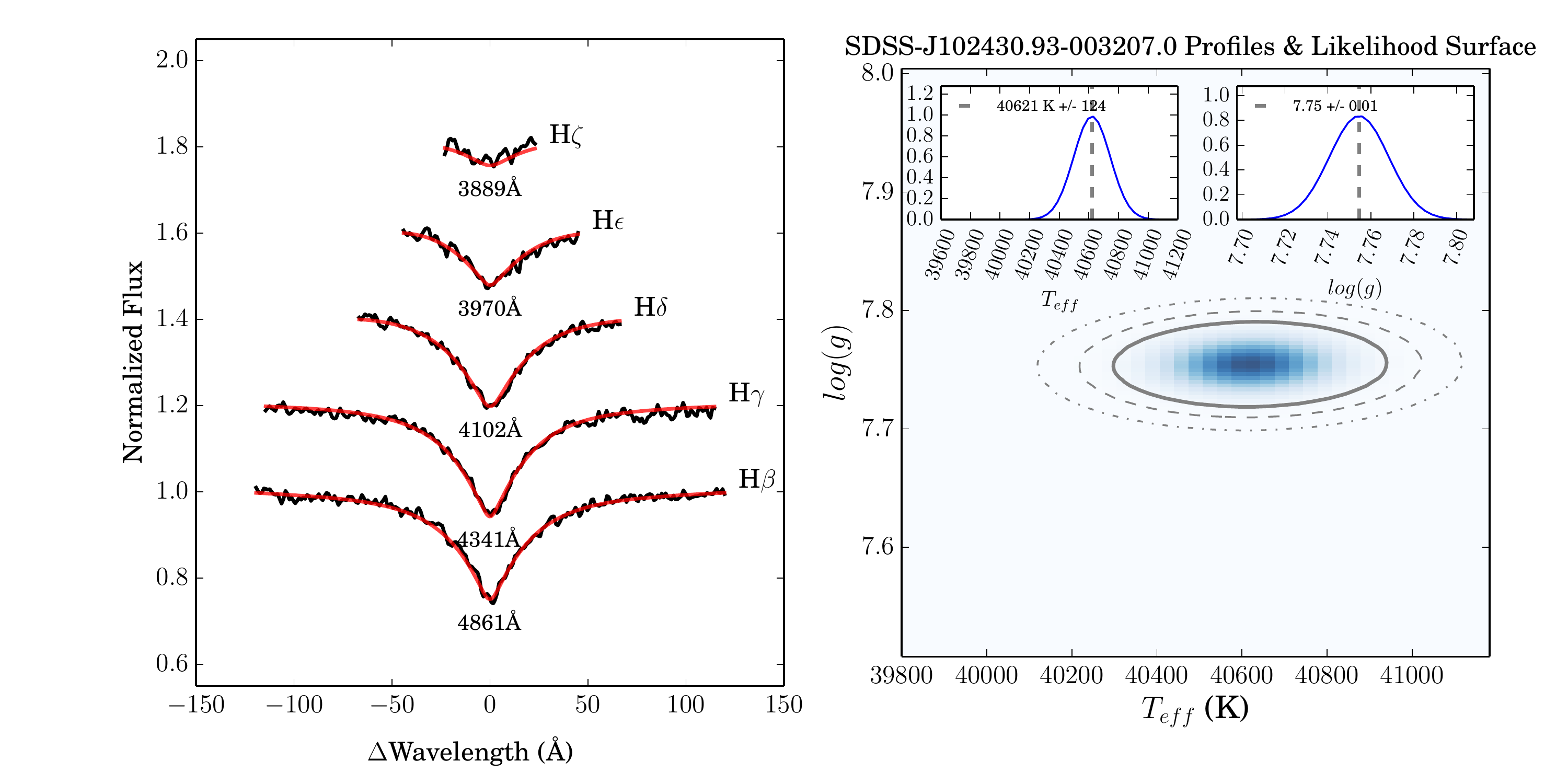}
\caption{(Left) Balmer line profiles extracted from the GMOS spectrum (black)
and model fits (red) for SDSS-J102430.93, the same object shown in
Figure~\ref{fig:spectra}.  The central wavelength of each feature is indicated
below the line. Small vertical offsets have been added to the continuum of each
line for visualization. Even the extremely weak H$\zeta$ feature is well
modeled.  (Right) The log likelihood surface in $T_{\text{eff}}$-$\log g$ space
for the fit are shown along with contours enclosing 68\%, 95\%, and 99.7\% of
the likelihood surface around the maximum likelihood estimate (solid, dashed,
and dot-dashed, respectively). Marginalizations are shown in the inset plots.
The 1-D distributions are well-modeled by Gaussians and the peak is indicated
in the legend.  There is virtually no covariance between the two parameters,
indicating that are observations are sufficient to constrain both
independently.}
\label{fig:prob}
\end{figure*}

We have analyzed the Balmer profiles of each star in a manner similar to that
described by \citet{Bergeron92} and \citet{Leibert05}. We use \texttt{Tlusty}
NLTE model atmospheres \citep{Hubeny95} (version
203)\footnote{\url{http://nova.astro.umd.edu/Tlusty2002/tlusty-frames-down.html}}
to match the models used by \citet{Bohlin14}. Each Balmer profile is
individually extracted with a window centered on the profile and extending well
into the wings. These extracted profiles are `flattened' by constructing a
linear interpolation between fixed points in the far wings of the profiles.
This flattening places the burden of determining $T_{\text{eff}}$ and $\log g$
solely on the quasi-symmetrical line profiles and decouples the analysis from
temperature dependent slopes in the underlying SED.  The model profiles are
flattened in precisely the same fashion.  Minimum $\chi^{2}$ fits are sought
between models and observations and contours enclosing 68\% and 95\% of the
total likelihood around the maximum likelihood estimate are constructed. The
profiles, fit, and likelihood surface for SDSS-J102430.93 are shown in
Figure~\ref{fig:prob}. 

The likelihood surfaces show very low covariance between $T_{\text{eff}}$ and
$\log g$. This is unsurprising as the two parameters change the shape of the
Balmer profiles in distinct ways.  Uncertainties are directly estimated from
the 1-D marginalized likelihoods for both parameters.  As the spectra have to
be dereddened before the line profiles are fit, we also examine the covariance
between $T_{\text{eff}}$ and $\log g$ and the fiducial value of $E(B-V)$ used
to deredden the spectrum, derived from SDSS DR12 photometry. We find a weak
linear covariance of both $T_{\text{eff}}$ and $\log g$ over a range of
$\pm0.05$ in $E(B-V)$. This range in $E(B-V)$ is much larger than the
uncertainty in the value determined from our WFC3 photometry.  We find that the
maximum difference in the best fit values for $T_{\text{eff}}$ and $\log g$
over the entire range of $E(B-V)$ are less than $0.3$ of the uncertainty in
$T_{\text{eff}}$ and $\log g$. The changes in $\chi^{2}$ over the same value
are not statistically significant and reflect the different centering of the
grid of $T_{\text{eff}}$ and $\log g$ over which the minimum chi-squared fit is
determined. This demonstrates that the temperature and surface gravity
determined from spectra are completely insensitive to the fiducial interstellar
$E(B-V)$ used to deredden the spectrum. 

\begin{figure*}[htb]
\centering
\epsscale{1.2}
\plotone{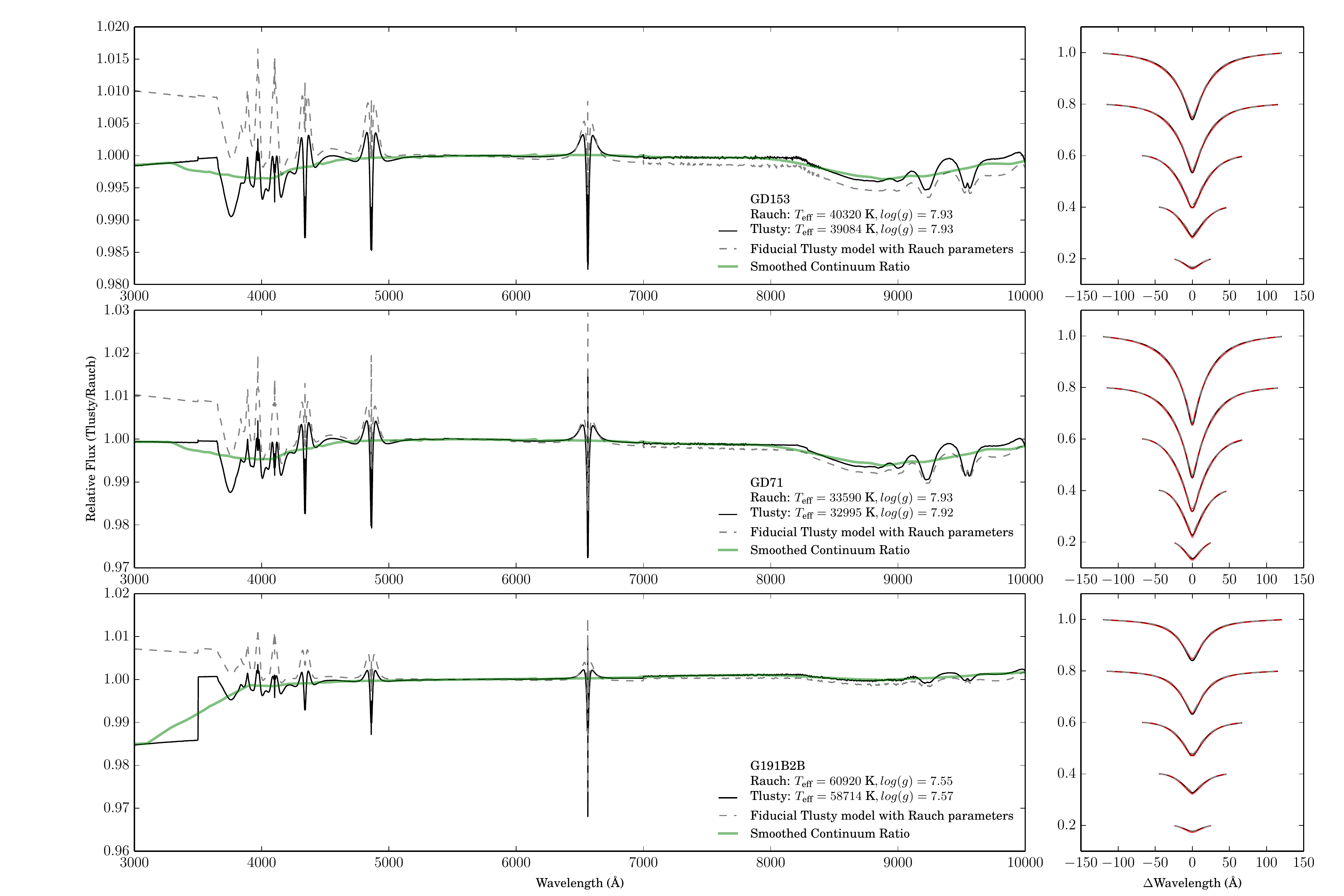}
\caption{Comparison of \texttt{Tlusty} \citep{Hubeny95} and \citet{Rauch13}
(RWBK) models of \emph {HST} primary spectrophotometric standards GD153 (top),
GD71 (middle) and G191B2B (bottom). The dashed gray line indicates the relative
flux ratio between the two models at the fiducial $T_{\text{eff}}$ and $\log g$
determined by \citet{Bohlin14b}. Fitting the line profiles of the RWBK models
using the \texttt{Tlusty} models improves the agreement between the two sets of
models (solid black lines). The residual differences are dominated by small
differences in the shapes of the line profiles between the two models, likely
because of different prescriptions used to model Stark broadening. Overall, the
line profile shapes are very similar (right panels). As these differences are
small and nearly symmetric about unity, they do not contribute significantly to
any net flux difference. We use a Savitzky-Golay filter to smooth the relative
flux ratio and highlight regions where the models disagree (solid green lines).
There are, however, no significant departures from unity.  Discontinuities in
the flux ratio in the UV arise from differences in the interpolation schemes
adopted by the two models and are only significant in the $F336W$ flux for
G191B2B. A change in the $T_{\text{eff}}$ of the \texttt{Tlusty} model of under
+1000K is sufficient to reduce the effect of the discontinuity to less than
$0.001$ in the relative flux ratio.  } \label{fig:modelcomp} \end{figure*}

Our methodology is insensitive to the choice of model. While model atmospheres
are generated as a function of $T_{\text{eff}}$ and $\log g$, these are
nuisance parameters that serve as labels for a specific line shape. Two
different models may have different values of $T_{\text{eff}}$ and $\log g$ for
the same line shape. Our procedure is only affected by \emph{relative} flux
differences at constant line shape. We estimate the systematic error arising
from the choice of a specific set of model atmospheres by fitting the
\citet[][hereafter RWBK]{Rauch13} models for the \emph{HST} primary standards
derived in \citet{Bohlin14b}. We fit the RWBK models using the same procedure
as the observed white dwarf spectra. The resulting model atmospheres are
normalized at 5556\AA.  The relative flux ratios are shown in Figure
\ref{fig:modelcomp}. 

The \texttt{Tlusty} models at the fiducial temperature and surface gravity of
the RWBK model are a close match to the \citet{Bohlin14b} models. Fitting the
line profiles of the RWBK models improves the agreement between the two models,
and the synthetic photometry difference is less than 0.003 mag in all
passbands. The residual difference in the line shape arises from a difference
in the shapes of the line profiles between the RWBK and \texttt{Tlusty} models,
likely a result of the different prescriptions used for Stark broadening. Sharp
discontinuities in the relative flux ratio arise from differences in the
continuum  between the two models. This is most evident in the models of
G191B2B. Both models are interpolated from model atmosphere grids, piece-wise
over different wavelength ranges, and these different sections are combined
together to create a model atmosphere spanning the full wavelength range at the
desired values of $T_{\text{eff}}$ and $\log g$. These discontinuities reflect
differences in the grids and the different interpolation schemes adopted by the
two models, rather than any difference in the underlying physics.

\subsection{Synthetic Photometry and Interstellar Reddening}\label{sec:synphotandred}

In most prior work on photometric standards, interstellar reddening has not
been a significant issue. The stars presented here, though, are from one to
three magnitudes fainter and significantly more distant so interstellar
reddening needs to be considered, estimated, and applied.  Our working model is
to assume that the DA white dwarf models correctly predict the emergent flux
from the star, and so the difference between observed magnitudes and model
predictions must show the color signature from the Galactic reddening law.
Since we have photometry in multiple bands, failure of either the Galactic
reddening law, or of the SED model for emergent flux, will show up as
discrepancies.  If all the passbands are consistent with these assumptions,
then the SED model plus the reddening/extinction value fully describes the
received SED from the white dwarf.  Accordingly we proceed as described below.

\subsubsection{Synthetic Photometry}

With the best fit temperature and surface gravity, we generate a full model SED
for the white dwarf, from 1300-25000\AA. We use the \texttt{pysynphot}
package,\footnote{\url{http://stsdas.stsci.edu/pysynphot/}} and convolve the
model SED with the response function of the observed \emph{HST} passbands. We
model the passband transmissions as the product of three components: the
optical train (Opt), the filter (PB) and the Quantum Efficiency of the WFC3
detectors (OE).

\begin{equation}
T(\lambda) = T_{\text{Opt}}(\lambda) \times T_{\text{PB}}(\lambda) \times \text{QE}(\lambda) 
\end{equation}

\noindent The published passband transmissions are based on pre-flight
laboratory measurements, and represent the fiducial system throughput. In
principle, it is possible to determine the count-rate yielded by an object
with an $F(\lambda)$ spectrum throughput by simply computing the integral of
the flux through the passband. However, all three components of the
transmission are time variable, and this is reflected in periodic updates to
the published MAST zeropoints. We therefore derive synthetic zeropoints for
each passband, $T$, using the definition of the \emph{HST} ``Vegamag'' system
that defines Vega to have a magnitude of zero at all wavelengths. The
synthetic flux, and magnitude, $\phi$, and $m$, respectively, is given by 

\begin{equation}
\phi^{S}_{T,i} = \frac{\int^{\infty}_{0} \lambda F^{S}_{i}(\lambda)T(\lambda) \,\mathrm{d}\lambda }{\int^{\infty}_{0} \lambda T(\lambda) \,\mathrm{d}\lambda}
\end{equation}

\begin{equation}
\phi^{S}_{T,\text{Vega}} =  \frac{\int^{\infty}_{0} \lambda F^{S}_{\text{Vega}}(\lambda)T(\lambda) \,\mathrm{d}\lambda }{\int^{\infty}_{0} \lambda T(\lambda) \,\mathrm{d}\lambda} 
\end{equation}

\begin{eqnarray}\label{eqn:synflux}
\begin{aligned}
m^{S}_{T,i} &= -2.5\log_{10}(\phi^{S}_{T,i}) + \text{ZP}^{\text{Vega}}_{T} \\
            &= -2.5\log_{10}(\phi^{S}_{T,i}) + 2.5\log_{10}(\phi^{S}_{T,\text{Vega}}) \\
%            &= -2.5\log_{10}(\phi^{S}_{T,i}) + 2.5\log_{10} \left( \frac{\int^{\infty}_{0} \lambda F^{S}_{\text{Vega}}(\lambda)T(\lambda)\,\,\mathrm{d}\lambda }{\int^{\infty}_{0} \lambda T(\lambda)\,\,\mathrm{d}\lambda} \right)  \\
            &= -2.5\log_{10} \left( \frac{\int^{\infty}_{0} \lambda F^{S}_{i}(\lambda)T(\lambda) \,\mathrm{d}\lambda }{\int^{\infty}_{0} \lambda F^{S}_{\text{Vega}}(\lambda) T(\lambda) \,\mathrm{d}\lambda} \right)  \\
\end{aligned}
\end{eqnarray}

\noindent where ``S'' is used to indicate synthetic quantities. The resulting
synthetic magnitudes are tied to Vega. The SED, $F^{S}(\lambda)$, is
not adjusted by the distance and radius of the white dwarf and thus is unreddened.
The reddening and normalization for the distance, and radius is treated in
\S\ref{sec:intred}.  We use the CALSPEC composite stellar spectrum of Vega
(\texttt{alpha\_lyr\_stis\_005})\footnote{\url{ftp://ftp.stsci.edu/cdbs/calspec/alpha\_lyr\_stis\_005.fits}}
as our model for Vega's SED, $F^{S}_{\text{Vega}}$. The CALSPEC model for Vega,
and GD153 used in \S\ref{sec:phot} are normalized consistently internally, and
thus synthetic magnitudes produced relative to the Vega model are directly
comparable to the observed magnitudes that have been tied to GD153. In Cycle
22, we will obtain observations of all three primary standards in all of our
bands, and tie our observations to the average of all three, obviating the need
to put our observations on the ``Vegamag'' system.

\subsubsection{Interstellar Reddening}\label{sec:intred}

We compute the difference between the observed and synthetic magnitudes for
each star, and fit these with i) a constant to account for the normalization
that is the same at all wavelengths (and reflects the radius of and distance to
the white dwarf), and ii) a slope corresponding to a scaling of the
\citet{Fitzpatrick99} reddening law, with the canonical $R_{V} = 3.1$,
appropriate for the Milky Way.  The extinction values in each passband are
given by the weighted extinction for each band as calculated via:

\begin{eqnarray}\label{eqn:synphot}
\begin{aligned}
m^{O}_{T,i} - m^{S}_{T,i} &= A_{T}(\left<\lambda\right>)*A^{i}_{V} + c_{i} \\
\left<\lambda\right> &= \frac{\int^{\infty}_{0} \lambda^{2} T(\lambda) \,\mathrm{d}\lambda }{\int^{\infty}_{0} \lambda T(\lambda) \,\mathrm{d}\lambda} \\
\end{aligned}
\end{eqnarray}
where $A(\left<\lambda\right>)$ is the extinction computed at the effective
wavelength of the passband $T$, $\left<\lambda\right>$, defined for unit
$E(B-V)$, again using the extinction of \citet{Fitzpatrick99} for $R_{V} =
3.1$, and $c_{i}$ is a wavelength-independent constant to account for a
passband independent overall brightness normalization for each white dwarf. The
normalization constant, $c_{i}$, must be added to the synthetic magnitudes,
$m^{S}_{T,i}$ to compute correctly normalized magnitudes.

\begin{figure}[htb]
\centering
\epsscale{1.2}
\plotone{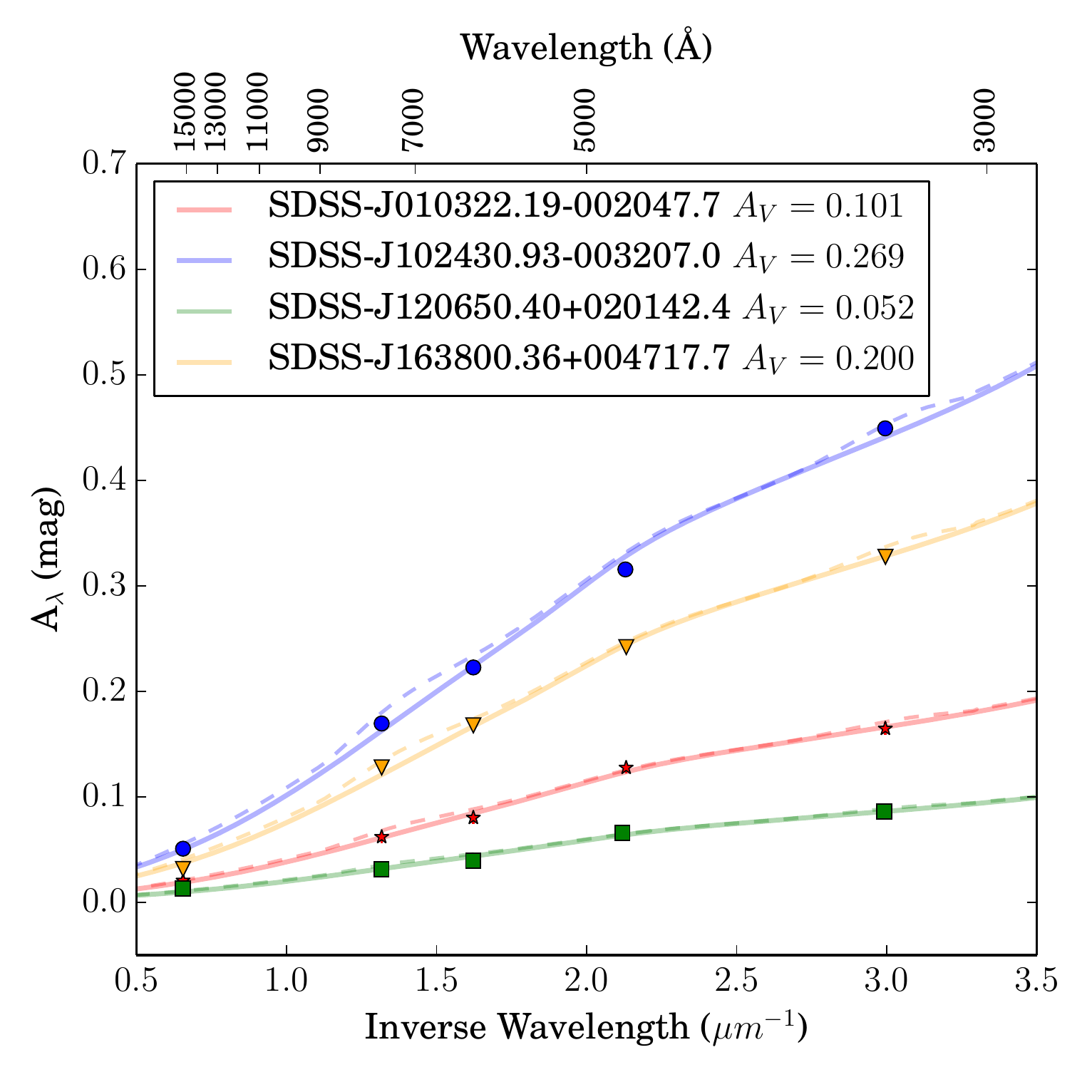}
\caption{Extinction curve fits with $R_{V} = 3.1$ using the reddening law of
\citet{Fitzpatrick99} (solid) and \citet{O'Donnell94} (dashed) for four objects
in our sample, spanning a range from low to high reddening. The largest
differences in the derived value of $A_{V}$ using the two different extinction
laws is seen for SDSS-J010322.19-002047.7 and SDSS-J120650.40+020142.4 where it
is 0.001 mag.  The mean residuals between the data and the fit extinction curve
are 3 mmag.}
\label{fig:extinction}
\end{figure}

The results of applying the above procedure to the four DA white dwarfs for
which the spectroscopic and photometric data are complete (discarding the
discrepant object with the emission components in the Balmer lines) are shown
in Table~\ref{tab:pho} and graphically in Figure~\ref{fig:extinction}.  Included
in the Table are the three fitted parameters, $A_{V}$, $T_{\text{eff}}$, and $\log g$,
intrinsic normalized magnitude before reddening, the extinction in each band,
and the net predicted magnitude per band. Both synthetic quantities include the
normalization constant, $c_{i}$. Comparison of column 9 with column 3
(predicted vs. observed magnitudes with WFC3/\emph{HST}) corresponds to the fit
residuals per band given in column 10.  Note that the extreme residuals are
$+/- 0.006$ with rms $ = 0.003$ mag.  

If we use the extinction law of \citet{O'Donnell94} instead of
\citet{Fitzpatrick99}, the derived $A_{V}$ values change by 1
milli-mag, thereby indicating that our experiment is insensitive to differences
between these forms of the reddening law. 

%the \citet{Fitzpatrick04} law fits with $R_{V} = 3.1$ are shown inFig.~\ref{fig:extinction}.

\section{Alternate Approach to Synthetic Photometry and Interstellar Reddening}
\label{sec:alt}

The above analysis is contingent on the state of calibration of WFC3,
and the assertion that CALSPEC standards have SEDs known very
accurately.  The shape of the WFC3 passbands is implicitly assumed to
be correct. The SED of GD153, a DA white dwarf itself, is used to
adjust the overall normalization of the effective passbands, and the
CALSPEC SED of Vega is used to establish the zeropoints of color.
These assumptions create a formal dependency on the state of
calibration of WFC3, and on the currently adopted SEDs of GD153 and
Vega.  The ``constructionist'' approach taken in the above section is
thus vulnerable to the state of calibration of WFC3.  Thus, we have
had to correct the zeropoints, as in equation \ref{eqn:zp1}.

However, there is indication that the  WFC3 zeropoints are uncertain at the
$0.01$~mag level.  For example, the current MAST zeropoints are prefaced by a
discussion that states ``The tables below summarize results determined from a
new reduction of all SMOV4, Cycle 17, and Cycle 18 observations of GD153,
G191B2B, GD71, and P330E. The independent calibrations from the four stars
agree to within 1\% in most filters, and the photometric zeropoint is set to
the average of the measurements.'' All four of these stars have nominally zero
reddening, and have equal weight in the determination of the WFC3 zeropoints. 

Given this reality, and our goal of achieving precisions substantially better
than 1\%, we have explored an alternate approach to optimally determining the
reddening and SEDs of our white dwarfs.  In this approach, we determine an
optimal set of stellar parameters and passband-dependent aperture corrections
by fitting the synthetic magnitudes for all stars simultaneously to the
observed instrumental magnitudes, including for each star an extinction value
and a passband-independent zeropoint.  The only assumptions here are that DA
white dwarf models do predict SEDs accurate to a few millimag or
better, 
the standard Galactic reddening law is valid for our targets, the
published WFC3 passbands are approximately correct in shape, and the bright DA
white dwarf GD153 has zero reddening.

We develop the following formalism, using the same notation adopted in \S~\ref{sec:synphotandred}.

The instrumental magnitude is just $-2.5 \log_{10} N$, where $N$ is the count rate of detected electrons.

\begin{equation}
N_{T,i} = k \int^{\infty}_{0} \lambda F^{O}_{i}(\lambda) T(\lambda) \,\mathrm{d}\lambda
\end{equation}

\noindent where $k$ is a \emph{band independent} constant that depends only on
the telescope aperture. We can write:

\begin{equation}
m^{X}_{T,i} = -2.5 \log_{10} k -2.5 \log_{10} \int^{\infty}_{0} \lambda F_{i}(\lambda) T(\lambda) \,\mathrm{d}\lambda
\end{equation}

We subsequently use $K = -2.5 \log_{10} k$. 

To compare a synthetic magnitude to a measured \emph{HST} instrumental magnitude, the
synthetic magnitude is defined in terms of the model flux, $F_{i}^{S}(\lambda)$ as

\begin{equation}
m^{S}_{T,i} = -2.5 \log_{10} \int^{\infty}_{0} \lambda F_{i}^{S}(\lambda) T(\lambda) \,\mathrm{d}\lambda
\end{equation}

We define another \emph{band independent} constant, $C_{i}$, to account for the
geometry (radius of the star, and distance to the star) that modulates for each
target star, the flux at the stellar surface to that incident at the top of the
terrestrial atmosphere. The instrumental magnitude, $m^{X}$, can then be
written in terms of synthetic ($S$) quantities as::

\begin{equation}
m^{X}_{T,i} = K -2.5 \log_{10} C^{X,S}_{i} -2.5 \log_{10} \int^{\infty}_{0} \lambda F_{i}^{S}(\lambda) T(\lambda) \,\mathrm{d}\lambda
\end{equation}

Denoting $K - 2.5 \log_{10} C^{X,S}_{i}$ by $ K^{X,S}_{i} $ this is simply:

\begin{equation}
m^{X}_{T,i} = K^{X,S}_{i} + m^{S}_{T,i}
\end{equation}

If the model fluxes are correct, the synthetic instrumental magnitudes will
equal the observed instrumental magnitudes. However, the synthetic magnitudes
implicitly count all photons detected from the object, regardless of where they
end up in the PSF, while the observed magnitudes count detected photons only
within a finite radius (0\farcs4 in our case).  The resulting difference is
the aperture correction and is band dependent because the PSF is.  A
nominal value of the aperture corrections is available from MAST, but we leave
them as values to be determined from the data. Denoting the aperture correction
as $\Delta_{T}$, we then expect that

\begin{equation}
m^{X}_{T,i}  = K^{X,S}_{i} + m^{S}_{T,i} + \Delta_{T}
\end{equation}

or, expressed another way,
\begin{equation}
\delta_{T,i} \equiv m^{X}_{T,i} - m^{S}_{T,i} - \Delta_{T} = K^{X,S}_{i}
\label{eqn:deltaij}
\end{equation}

Note that, although we expect the dominant contribution to $ \Delta_{T} $ to
come from the aperture correction, \textit{any} bandpass dependent error,
including putative errors in calibrating the instrument,  will contribute.  The
dependence of $\delta_{T,i}$ on the passband, $T$, is a powerful diagnostic.
Equation \ref{eqn:deltaij} tells us that the wavelength dependence should be
fully accounted for by the aperture correction, and the synthetic model.  If
there is a discrepancy, it can arise from several causes:
\begin{itemize}
\item The aperture corrections $\Delta_{T}$ are incorrect.
\item There is extinction affecting the real object that is not represented in the model.
\item The model fluxes have errors not accounted for by extinction.
\end{itemize}

Given the distance to our stars, it is likely that they will be affected by
some amount of extinction (interstellar, and possibly circumstellar), we can
fit for the reddening, and analyze the remaining residuals for the other causes listed
above.  The extinguished synthetic magnitude is:
\begin{equation}
m^{S}_{T,i} = -2.5 \log_{10} \int^{\infty}_{0} \lambda F^{S}_{i}(\lambda) \alpha_{i}(\lambda) T(\lambda) \,\mathrm{d}\lambda
\end{equation}

\noindent where $\alpha_{i}(\lambda)$ is the transmission function of the extinction for object $i$.

We parameterize $\alpha_{i}(\lambda)$ in terms of the quantities $R_V$ and
$E_{(B-V,i)}$, and calculate it by the prescription of \citet{O'Donnell94}. We
can then determine the extinction parameters, along with the $\Delta_{T}$ and
$K^{X,S}_{i}$ by minimizing

\begin{eqnarray}\label{eqn:chisq}
\begin{aligned}
\chi^2&(\{E_{B-V,i}\}, \{K^{X,S}_{i}\}, \{\Delta_T\}) = \\
\sum_{T,i} &(m^{X}_{T,i} - m^{S}_{T,i} \\
         - &A_{T}\times R_{V}\times E_{i}(B-V) \\
         - &\Delta_T - K^{X,S}_{i})^2 \\
\end{aligned}
\end{eqnarray}

\noindent with the constraint that $E_{B-V,i} \geq 0 ~ \forall i$.

With our current four stars and 5 WFC3 passbands, $\chi^2$ depends on 12
parameters, while we have 20 measurements, so the minimization problem is
potentially well constrained.  Additional stars and/or passbands will constrain
the solution further.  We keep $R_V = 3.1$ for the present, although it could
in principle be included in the parameters being optimized.

There are two difficulties with this approach, however.  The first one is
trivial.  There is a partial degeneracy between the $\Delta_{T}$ and
$K^{X,S}_{i}$, in that an arbitrary constant can be added to all the
$\Delta_{T}$ and then subtracted from all the $K^{X,S}_{i}$ without changing
the value of $\chi^2$.  This needlessly complicates the job of the optimizer,
and makes the results harder to interpret.  We deal with this by enforcing

\begin{equation}
\sum_T \Delta_T = 0
\end{equation}

The second difficulty is more serious. Experience with minimizing $\chi^2$ as
given above shows that the minimum is consistently achieved by choosing the
$\Delta_{T}$ so that the extinction of the least extinguished star comes out to be
zero.  This is readily achieved, since the right choice for $\Delta_{T}$ can
mimic an arbitrary extinction, that is then effectively subtracted from the
extinctions of all the stars.  This behavior is unphysical, and must be
remedied by an additional data-based constraint. We have chosen to use the
observations of GD153 discussed in Section \ref{sec:phot} in conjunction with
the conclusion of \cite{Bohlin14} that its extinction is zero to high
precision.  

To make use of this constraint, we treat the GD153 data in the same way as the
program stars, but its contribution to $\chi^2$ is treated differently in that
when calculating the model flux, $E_{B-V,\text{GD153}} = 0$ by assumption.  The GD153
contribution to the overall $ \chi^2 $ is multiplied by a weighting factor
large enough to ensure that any significant deviation from this assumption is
suppressed. The summation over bands includes only $F336W$, $F475W$, and $F625W$,
since we currently lack GD153 data for the other bands.  The model flux for
GD153 uses the G11 values for $T_{\text{eff}}$ and $\log g$ from
\cite{Bohlin14b}. 

\begin{equation}
\chi_{\text{GD153}}^2 = w_{\text{GD153}} \sum_i (m^{X}_{T,\text{GD153}} - m^{S}_{T,\text{GD153}} -\Delta_T-K^{X,S}_{\text{GD153}})^2
\end{equation}
The weighting factor $w_{\text{GD153}} = 1/0.01^2$ for the results quoted below ensures that the GD153 reddening is very close to zero.  

\subsection{Results}
The $\chi^2$ minimization of equation \eqref{eqn:chisq} was carried out using
IDL routine \verb|CONSTRAINED_MIN|.  To perform error analysis of the results,
the minimization was carried out a large number of times using independent
samples from Gaussian distributions of $T_{\text{eff}}, \log g$ and
$\{m^{O}_{T,i}\}$.  The covariance matrix for these quantities is assumed to be
diagonal.  The uncertainties in the distributions come from Table
\ref{tab:pho}.  Table~\ref{tab:extinct} shows the resulting mean values of
$A_V$ for the four stars, their statistical uncertainties, and the mean
standard deviation of the fit to the observed magnitudes for each star.
Table~\ref{tab:Deltashift} presents the mean values of the derived aperture
corrections, $\Delta_{T}$, for the five WFC3 bands we employ, and the nominal
values from MAST.  It is noteworthy that the agreement between our derived
aperture corrections, $\Delta_{T}$, and the fiducial MAST values is excellent,
with discrepancies of order 0.01 mag, as expected.

Comparison with $A_{V}$ values in Table~\ref{tab:extinct} with those in
Table~\ref{tab:pho} that are  derived by the other procedure shows differences
of a few millimag, with consequent differences in SED fluxes over the observed
passband range of 10 or 20 millimag.  We expect that with the Cycle 22 data,
which includes observation in $F275W$ and contemporaneous constraining
observation of all three \emph{HST} white dwarf calibrators, the SEDs from both
approaches will result in closer agreement for the derived final SEDs.  We
emphasize that all quoted uncertainties are purely statistical.

\begin{deluxetable}{lccc}
\tabletypesize{\scriptsize}
\tablewidth{0pt}
\tablecolumns{4}
\tablecaption{Fit Results for Primary DA White Dwarf Targets\label{tab:extinct}}
\tablehead{
    \colhead{Object} &
    \colhead{$A_V$} &
    \colhead{$\sigma_{A_V}$}                                  &
    \colhead{$\sigma(fit)$} \\
    \colhead{} &
    \multicolumn{3}{c}{(mag)}                                      
     }
\startdata
SDSS-J010322.19-002047.7 &  0.095   & 0.005 & 0.007  \\
SDSS-J102430.93-003207.0 &  0.264   & 0.005 & 0.004   \\
SDSS-J120650.40+020142.4 &  0.051   & 0.008 & 0.004 \\
SDSS-J163800.36+004717.7 &  0.197   & 0.005 & 0.006 \\
\enddata
\end{deluxetable}

\begin{deluxetable*}{lrr}
\tabletypesize{\scriptsize}
\tablewidth{0pt}
\tablecolumns{3}
\tablecaption{Derived and Fiducial Aperture Corrections for \emph{HST} passbands\label{tab:Deltashift}}
\tablehead{
    \colhead{Passband} &
    \colhead{Aperture Correction} &
    \colhead{Mean subtracted} \\                                       
    \colhead{} &
    \colhead{$\Delta_{T}$} & 
    \colhead{MAST Aperture Correction} \\
    \colhead{} &
    \multicolumn{2}{c}{(mag)} 
     }
\startdata
F336W &     0.011    &    0.033  \\
F475W &  $-$0.039    & $-$0.023  \\
F625W &  $-$0.023    & $-$0.024  \\
F775W &  $-$0.095    & $-$0.020  \\
F160W &     0.060    &    0.064  \\
\enddata
\end{deluxetable*}

\section{Expected Magnitudes on Common Photometric Systems}
\label{sec:predicts}

As our targets are within the dynamic range of present wide-field survey
facilities, we provide expected synthetic magnitudes for SDSS, PanSTARRS, and
DECam photometric systems in Table~\ref{tab:extraphot}.  The passband
throughputs used to synthesize magnitudes include the attenuation from the
terrestrial atmosphere at a representative airmass of $1.3$ at the respective
observatory sites. Where available, we have compared our synthetic photometry
to published photometry from surveys from their respective facilities.

In all cases we quote magnitudes on the AB system, as defined for photon
proportional detection systems in \cite{Fukugita96} (see their equation 7). We
have converted from our \emph{HST}-based photometric system to AB by
determining the AB magnitude of Vega. 

The mean absolute value of the residuals between the observed and synthetic
photometry are within $0.025$~mag for the $ugri$ bands within SDSS DR12,
completely in line with the 2--3\% systematic error associated with the
absolute calibration of the SDSS photometry onto the AB system.

The observed PanSTARRS photometry is from \citet{Tonry12} and
\citet{schlafly12}, and adjusted by the offsets reported in \citet{Scolnic15}.
There are no measurements reported for SDSSJ102430.93. The residuals to the
PanSTARRS photometry show a negative bias in the $y$ band, however this bias is
within $2--3\sigma$ of the large reported uncertainties of the photometry. The
PanSTARRS 3pi survey images have significantly more cosmetic issues than the
SDSS photometry, and careful analysis of these images with imporoved masking of
diffraction spikes, ghosts, and cosmic rays will likely improve the agreement
between the observed and synthetic photometry.

\begin{deluxetable*}{clcllllll}
\tabletypesize{\scriptsize}
\tablewidth{0pt}
\tablecolumns{8}
\tablecaption{DA White Dwarf Synthetic and Observed Magnitudes on Common Photometric Systems\label{tab:extraphot}}
\tablehead{
    \colhead{Survey} &
    \colhead{Target} &
    \colhead{Source\tablenotemark{a}} &
    \colhead{\emph{u}} &
    \colhead{\emph{g}} &
    \colhead{\emph{r}} &
    \colhead{\emph{i}} &
    \colhead{\emph{z}} &
    \colhead{\emph{y}} \\
    \colhead{} &
    \colhead{} &
    \colhead{} &
    \multicolumn{6}{c}{(mag)\tablenotemark{b}}
}
\startdata
DECam &    SDSS-J010322.19 & Syn &  18.744 & 19.096 & 19.629 & 19.999 & 20.293 & 20.474 \\
      &    SDSS-J102430.93 & Syn &  18.661 & 18.914 & 19.365 & 19.694 & 19.956 & 20.127 \\
      &    SDSS-J120650.40 & Syn &  18.630 & 18.683 & 19.115 & 19.444 & 19.721 & 19.879 \\
      &    SDSS-J163800.36 & Syn &  18.499 & 18.831 & 19.332 & 19.683 & 19.961 & 20.137 \\
\\
PS1\tablenotemark{c}   &    SDSS-J010322.19 & Syn &         & 19.109         & 19.573         & 19.938         & 20.212         & 20.484         \\
      &                    & Obs &         & 19.104 (0.007) & 19.432 (0.016) & 19.944 (0.021) & 20.132 (0.032) & 20.120 (0.124) \\
      &                    & Res &         & $-$0.005       & $-$0.141       & $+$0.006       & $-$0.080       & $-$0.364       \\
\\
      &    SDSS-J102430.93\tablenotemark{d} & Syn &         & 18.925         & 19.315         & 19.642         & 19.886         & 20.136         \\
\\
      &    SDSS-J120650.40 & Syn &         & 18.689         & 19.065         & 19.386         & 19.643         & 19.888         \\
      &                    & Obs &         & 18.688 (0.007) & 19.030 (0.010) & 19.375 (0.014) & 19.645 (0.026) & 19.742 (0.052) \\
      &                    & Res &         & $-$0.001       & $-$0.035       & $-$0.011       & $+$0.002       & $-$0.146       \\
\\
      &    SDSS-J163800.36 & Syn &         & 18.843         & 19.279         & 19.628         & 19.886         & 20.147         \\
      &                    & Obs &         & 18.879 (0.008) & 19.281 (0.010) & 19.610 (0.011) & 19.880 (0.021) & 19.963 (0.101) \\
      &                    & Res &         & $+$0.036       & $+$0.002       & $-$0.018       & $-$0.006       & $-$0.184       \\
\\
SDSS  &    SDSS-J010322.19 & Syn &  18.657        & 19.045        & 19.553        &  19.918       & 20.255       \\
      &                    & Obs &  18.66 (0.02)  & 19.03 (0.01)  & 19.54 (0.01)  &  19.94 (0.02) & 20.37 (0.12) \\
      &                    & Res &  $~$0.00          & $-$0.02       & $-$0.01       & $+$0.02    & $+$0.11         \\
\\ 
      &    SDSS-J102430.93 & Syn &  18.597        & 18.874        & 19.296        &  19.623       & 19.922       \\
      &                    & Obs &  18.57 (0.02)  & 18.84 (0.01)  & 19.28 (0.01)  &  19.58 (0.02) & 19.77 (0.08) \\
      &                    & Res &  $-$0.03       & $-$0.03       & $-$0.02       & $-$0.04       & $-$0.15      \\
\\
      &    SDSS-J120650.40 & Syn &  18.586        & 18.648        & 19.046        &  19.367       & 19.681       \\
      &                    & Obs &  18.55 (0.02)  & 18.63 (0.01)  & 19.03 (0.01)  &  19.34 (0.02) & 19.70 (0.10) \\
      &                    & Res &  $-$0.04       & $-$0.02       & $-$0.02       & $-$0.03       & $+$0.02         \\
\\
      &    SDSS-J163800.36 & Syn &  18.418        & 18.783        & 19.259        &  19.608       & 19.925       \\
      &                    & Obs &  18.44 (0.01)  & 18.81 (0.01)  & 19.28 (0.01)  &  19.58 (0.02) & 19.79 (0.08) \\
      &                    & Res &  $+$0.02          & $+$0.03          & $+$0.02          & $-$0.03       & $-$0.14      \\
\enddata
\tablenotetext{a}{Synthetic magnitudes are labeled ``Syn''. Observed magnitudes are provided where available and are labeled ``Obs''. Uncertrainties are provided in parentheses. Residuals between Observed and synthetic magnitudes are labeled by ``Res''.}
\tablenotetext{b}{All magnitudes are on the AB system. The transmission/response function for each survey facility is calculated assuming an airmass of 1.3.}
\tablenotetext{c}{The PS1 observed magnitudes are from the calibration of \citet{Tonry12} and \citet{schlafly12}. They have been adjusted by the offsets in \citet{Scolnic15}. These offsets are $(\Delta g, \Delta r, \Delta i, \Delta z, \Delta y) = (0.020, 0.033, 0.024, 0.028, 0.011)$ mag.}
\tablenotetext{d}{There are no entries for SDSS-J102430.93 in the catalogs of \citet{Tonry12} and \citet{schlafly12}.}
\end{deluxetable*}

\section{Conclusions and Discussion}
\label{sec:Concl}

We have presented the results from four faint DA white dwarfs that
demonstrate the proposition that DA white dwarf fluxes as predicted
from models are consistent with observations.  This proposition has
been the basis for \emph{HST} calibrated fluxes for some time now.  In
this paper we have demonstrated that for fainter objects, where modest
amounts of reddening must taken into account, the simple extension
that includes solving for and applying individual self-determined
reddening corrections over a wide range of wavelengths using a standard
Milky Way reddening law works quite well, producing residuals of a few
millimag.  Nevertheless, there are several caveats to consider, some
of which will be addressed in our continuing work as more data on more
objects become available.

What our results really show is that the models accurately predict SED
{\it ratios} between DA white dwarfs, i.e., the models are validated as
{\it differential} predictors.  It does not, however, test whether the
models predict the absolute SEDs correctly. This is likely the dominant source 
of systematic uncertainty.  We can get some purchase on it by comparing different
sets of DA white dwarf models. A comprehensive discussion can be found  
in \citet{Bohlin14}, where current models are found to differ by more than 1\% 
(and as much as 5\%) short-ward of
$0.27 ~\mu {\rm m}$  and long-ward of $5 ~\mu {\rm m}$
It is also possible that our understanding of
the relevant atomic and atmospheric processes is yet
incomplete. Nevertheless the validation of models as a good
differential predictor of SEDs is already a tool that will enable us
to calibrate photometric systems to a few millimag accuracy.  The data
we present here (and in future papers to come) will continue to anchor
results, as model improvements and validation take place.

There are also additional caveats, but for these we have specific mitigation plans:

\begin{enumerate}
\item
There are improvements underway to our processing steps. In
particular, there is covariance between effective temperature and
reddening. The data analysis flow is being redesigned to fit both
these parameters (along with $\log g$) in the same process.
\item
For the 9 targets observed in the \emph{HST} GO-12967 (Cycle
20) program, there are additional photometric data being gathered in
the Cycle 22 \emph{HST} GO-13711 program.  The additional images will
not only secure the photometry, but an additional passband (F275W in
the near UV) will allow us to examine if the standard extinction law
for reddening is indeed adequate.
\item
We are obtaining contemporaneous observations of all three \emph{HST} primary
standards, GD153, GD71 and G191B2B in our Cycle 22 program. This will allow us
to tie our measurements directly on to the CALSPEC photometric scale, without
the additional step of tying to the ``Vegamag'' scale, that in turn is tied
to CALSPEC.
\item
There has been little or no long term monitoring of these objects for
constancy.  We are seeking to pursue this from the ground.
\item 
It is known that a small fraction of DA white dwarfs have gaseous and planetary
debris around them and some display pollution from metals \citep[e.g.,
][]{koester14}.  These can produce anomalous SEDs.  Metal pollution is
detectable only in the far UV.  A program to observe our target DA white dwarfs
with COS on \emph{HST} would be necessary to constrain this potential
systematic.
\item
Magnetic fields have sometimes been observed in DA stars in our temperature
range.   The effect of such fields can be two fold.  First, undetected Zeeman
splitting in the Balmer lines can bias temperature and gravity determinations.
Second,  a rotating magnetic white dwarf can exhibit observable flux variations
\citep{Holberg12}.  Flux monitoring and spectroscopy of the H$\alpha$ line can
flag the presence of such effects.
\end{enumerate}

As a result of the above cautionary points, and with specific plans
under way for their mitigation, we urge that the results for the four
DA white dwarfs presented in Tables~\ref{tab:pho} and
\ref{tab:extraphot} be treated as provisional.  Nevertheless, we will
be very surprised if they change by more than a percent and we
publish them so that they can be tested by observers.

%%%%%%%%%%%%%%%%%%%%%%%%%%%%%%%%%%%%%%%%%%%%%%%%%%%%%%%%%%%%%%%%%%%%%%%%%%%%%%%%%%%
\begin{acknowledgments}

Some of the data presented in this paper were obtained from the
Mikulski Archive for Space Telescopes (MAST). STScI is operated by the
Association of Universities for Research in Astronomy, Inc., under
NASA contract NAS5-26555. Support for MAST for non-\emph{HST} data is
provided by the NASA Office of Space Science via grant NNX13AC07G and
by other grants and contracts.

Based on observations obtained at the Gemini Observatory, which is
operated by the Association of Universities for Research in Astronomy,
Inc., under a cooperative agreement with the NSF on behalf of the
Gemini partnership: the National Science Foundation (United States),
the National Research Council (Canada), CONICYT (Chile), the
Australian Research Council (Australia), Minist\'{e}rio da
Ci\^{e}ncia, Tecnologia e Inova\c{c}\~{a}o (Brazil) and Ministerio de
Ciencia, Tecnolog\'{i}a e Innovaci\'{o}n Productiva (Argentina)
(Programs GS-2013A-Q-8 and GS-2013B-Q-22).

The authors thank the anonymous referee for incisive questions and helpful
suggestions.

The authors wish to thank Dr. Annalisa Calamida for useful discussion and her
critical reading of the paper.

{\it Facilities:}
\facility{\emph{HST} (WFC3, ACS)}, 
\facility{Gemini:South (GMOS)}, 
\facility{MMT (Blue Channel)}, 
\facility{Magellan:Baade (IMACS)}. 
\end{acknowledgments}

%%%%%%%%%%%%%%%%%%%%%%%%%%%%%%%%%%%%%%%%%%%%%%%%%%%%%%%%%%%%%%%%%%%%%%%%%%%%%%%%%%%

\bibliographystyle{apj}
\bibliography{wd}

%%%%%%%%%%%%%%%%%%%%%%%%%%%%%%%%%%%%%%%%%%%%%%%%%%%%%%%%%%%%%%%%%%%%%%%%%%%%%%%%%%%

\end{document}